\documentclass[preprint,3p,11pt,sort&compress]{elsarticle}

\usepackage{booktabs}
\usepackage{amsmath}
\usepackage{amssymb}
\usepackage{lineno}
\usepackage{color}
\usepackage{url}
\usepackage{bm}

\journal{arXiv}

\begin{document}

% \linenumbers

\begin{frontmatter}

\noindent \textcolor{blue}{This manuscript has been accepted for publication in \textit{Structural Safety} and is already published.
The DOI is \url{https://doi.org/10.1016/j.strusafe.2024.102442}.
Please cite this as follows. T. Yaoyama, T. Itoi, J. Iyama, Probabilistic Model Updating of Steel Frame Structures Using Strain and Acceleration Measurements:
A Multitask Learning Framework, Structural Safety 108 (2024) 102442.}

\title{Probabilistic Model Updating of Steel Frame Structures Using Strain and Acceleration Measurements: A Multitask Learning Framework}

\author[label1]{Taro Yaoyama\corref{cor1}}\ead{yaoyama@g.ecc.u-tokyo.ac.jp}
\author[label1]{Tatsuya Itoi}
\author[label1]{Jun Iyama}

\cortext[cor1]{corresponding author}

\affiliation[label1]{
    organization={Department of Architecture, Graduate School of Engineering, The University of Tokyo},
    addressline={7-3-1, Hongo}, 
    city={Bunkyo-Ku},
    postcode={113-8656}, 
    state={Tokyo},
    country={Japan}
}

\begin{abstract}

This paper proposes a multitask learning framework for probabilistic model updating by jointly using strain and acceleration measurements.
This framework can enhance the structural damage assessment and response prediction of existing steel frame structures with quantified uncertainty.
Multitask learning may be used to address multiple similar inference tasks simultaneously to achieve a more robust prediction performance by transferring useful knowledge from one task to another, even in situations of data scarcity.
In the proposed model-updating procedure, a spatial frame is decomposed into multiple planar frames that are viewed as multiple tasks and jointly analyzed based on the hierarchical Bayesian model, leading to robust estimation results.
The procedure uses a displacement--stress relationship in the modal space because it directly reflects the elemental stiffness and requires no prior knowledge concerning the mass, unlike most existing model-updating techniques.
Validation of the proposed framework by using a full-scale vibration test on a one-story, one-bay by one-bay moment resisting steel frame, wherein structural damage to the column bases is simulated by loosening the anchor bolts, is presented.
The experimental results suggest that the displacement--stress relationship has sufficient sensitivity toward localized damage, and the Bayesian multitask learning approach may result in the efficient use of measurements such that the uncertainty involved in model parameter estimation is reduced.
The proposed framework facilitates more robust and informative model updating.

\end{abstract}

% % Graphical abstract
% \begin{graphicalabstract}
% % \includegraphics{grabs}
% \end{graphicalabstract}

% \begin{highlights}
%     \item Research highlight 1
%     \item Research highlight 2
% \end{highlights}


\end{frontmatter}

\section{Introduction}

Finite element (FE) model updating techniques calibrate the parameters of analytical models such that their outputs correlate with the measurements of in-service structures and enhance structural damage detection and response prediction for future potential hazards such as earthquakes.
However, these techniques suffer from ill-posed and ill-conditioned problems attributed to measurement noise and modeling errors originating from multiple sources\cite{mottersheadJ2011,simoenE2015}.
Thus far, Bayesian approaches to FE model updating have been investigated by several researchers\cite{beckJL1998,katafygiotisLS1998,beckJL2002,chingJ2007,beckJL2010,straubD2015,rocchettaR2018} because they provide a rigorous framework to incorporate uncertainty quantification (UQ) in the model-updating procedure.

As a more effective countermeasure for ill-posed and ill-conditioned inverse problems, the hierarchical Bayesian model (HBM) has received increasing interest in the context of model updating and structural health monitoring (SHM)\cite{behmaneshI2015,songM2019,sadehiO2019,jiaX2022,huangY2017a,huangY2017b,houR2019,filippitzisF2022}.
Behmanesh et al.\cite{behmaneshI2015} presented an HBM framework that flexibly quantified the inherent variability of structural parameters because of environmental variations (such as changing temperature) by setting hierarchical prior distributions over these parameters.
Similar formulations were made for quantifying the variability attributed to the excitation amplitude\cite{songM2019}, time-domain model updating\cite{sadehiO2019}, and hysteretic model updating\cite{jiaX2022}.
Sparse Bayesian learning (SBL) for model updating purposes was formulated based on the HBM\cite{huangY2017a,huangY2017b,houR2019,filippitzisF2022}, where the sparsity of damage locations in a structure was assumed in the absence of structural collapse.

Hierarchical modeling serves to develop a multitask learning framework called Bayesian multitask learning. Multitask learning addresses multiple similar inference tasks jointly by transferring useful knowledge from one task to another and achieves a more robust prediction performance, especially in situations where data sparsity is crucial if multiple tasks are individually tackled\cite{zhangY2018}.
Although multitask learning has several SHM applications\cite{wanHP2018,huangY2019a,huangY2019c,gardnerP2021,gardnerP2022,bullLA2022}, its application to Bayesian model updating remains limited.
In a pioneering study, Huang et al.\cite{huangY2019a} extended the aforementioned studies\cite{huangY2017a,huangY2017b} to a multitask learning approach, where damage locations in different datasets were expected to have similar sparseness profiles.

This study entails the development of a multitask learning framework for probabilistic model updating based on HBM that specifically focuses on the joint use of strain and acceleration measurements. Many studies on model updating have used modal properties (e.g., natural frequencies, damping ratios, and displacement mode shapes) or frequency-domain properties (e.g., frequency response function or power spectral density), both of which are extracted from acceleration measurements. However, acceleration represents the characteristics of an entire structure instead of those of its individual elements. From this viewpoint, dynamic strain measurements may be more useful to update the FE model because of their higher sensitivity to elemental stiffness, as demonstrated in the literature\cite{esfandiariA2010,pedramM2016,singhMP2018,matarazzoTJ2018}.
The combined use of strain and acceleration measurements was investigated in several studies \cite{ungerJF2005,iyamaJ2021,iyamaJ2023}.
Iyama et al.\cite{iyamaJ2021,iyamaJ2023} used both strain and displacement mode shapes for computing the relationship between the displacement and locally distributed stress (axial force and bending moment), which directly provided useful insights into the elemental stiffness and effectively allowed damage localization and quantification.
Such a stress-based procedure required no prior knowledge of the mass, whereas many model-updating studies assumed that the mass is known in advance.

This paper presents a probabilistic method for the stress-based model updating of steel frame structures. A spatial frame is decomposed into multiple planar frames viewed as multiple tasks and jointly inferred in a multitask framework for more efficient use of measurements.
The main contributions of this study can be summarized as follows:
\begin{enumerate}
    \renewcommand{\labelenumi}{(\arabic{enumi})}
    \item It provided an overview of the procedure for relating displacement to elemental stress (especially bending moment) based on both strain and displacement mode shapes extracted through system identification techniques, which was presented by Iyama et al.\cite{iyamaJ2021,iyamaJ2023} and Yaoyama et al.\cite{yaoyamaT2023}.
    \item It entailed the formulation of a Bayesian multitask learning framework for model updating based on such a displacement--stress relationship, termed the \textit{normalized modal bending moment}, which requires no prior knowledge about mass, unlike most existing model-updating techniques.
    \item It entailed an experimental study using a full-scale steel frame structure to demonstrate that the proposed method is sufficiently sensitive for detecting localized stiffness decreases and that the multitask learning framework reasonably reduces uncertainty in parameter estimation and response prediction.
\end{enumerate}

The remainder of this paper is organized as follows.
Section 2 describes the experimental setup and data pre-processing. Section 3 presents the proposed framework and formulates its application to the target experimental data. Section 4 presents the results of system identification and model updating and discusses the effectiveness of the proposed framework. Finally, Section 5 presents the main conclusions and suggestions for future research.

\begin{figure}[!t]
    \centering
    \includegraphics[width=80truemm]{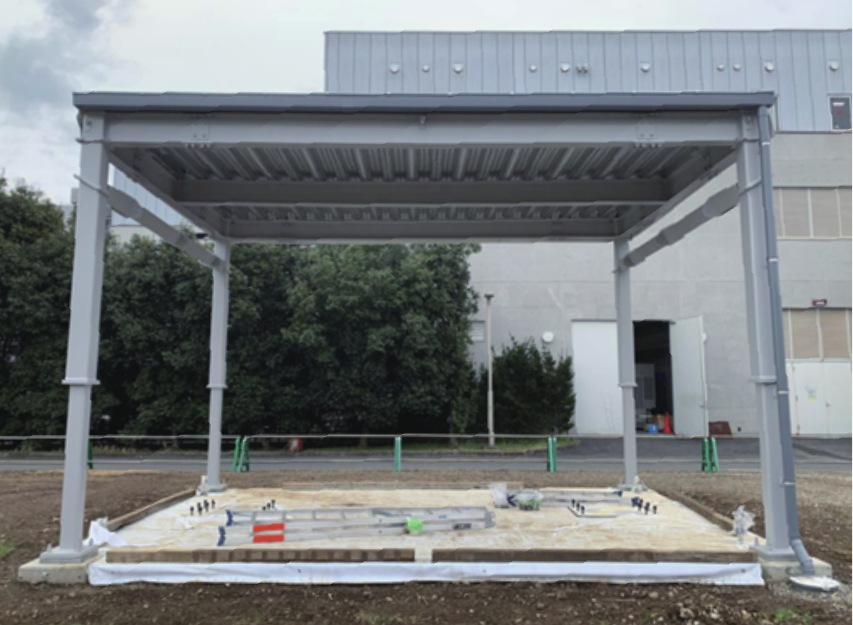}
    \caption{Experimental model.}\label{fig:photo}
\end{figure}

\begin{figure}[!t]
    \centering
    \includegraphics[width=150truemm]{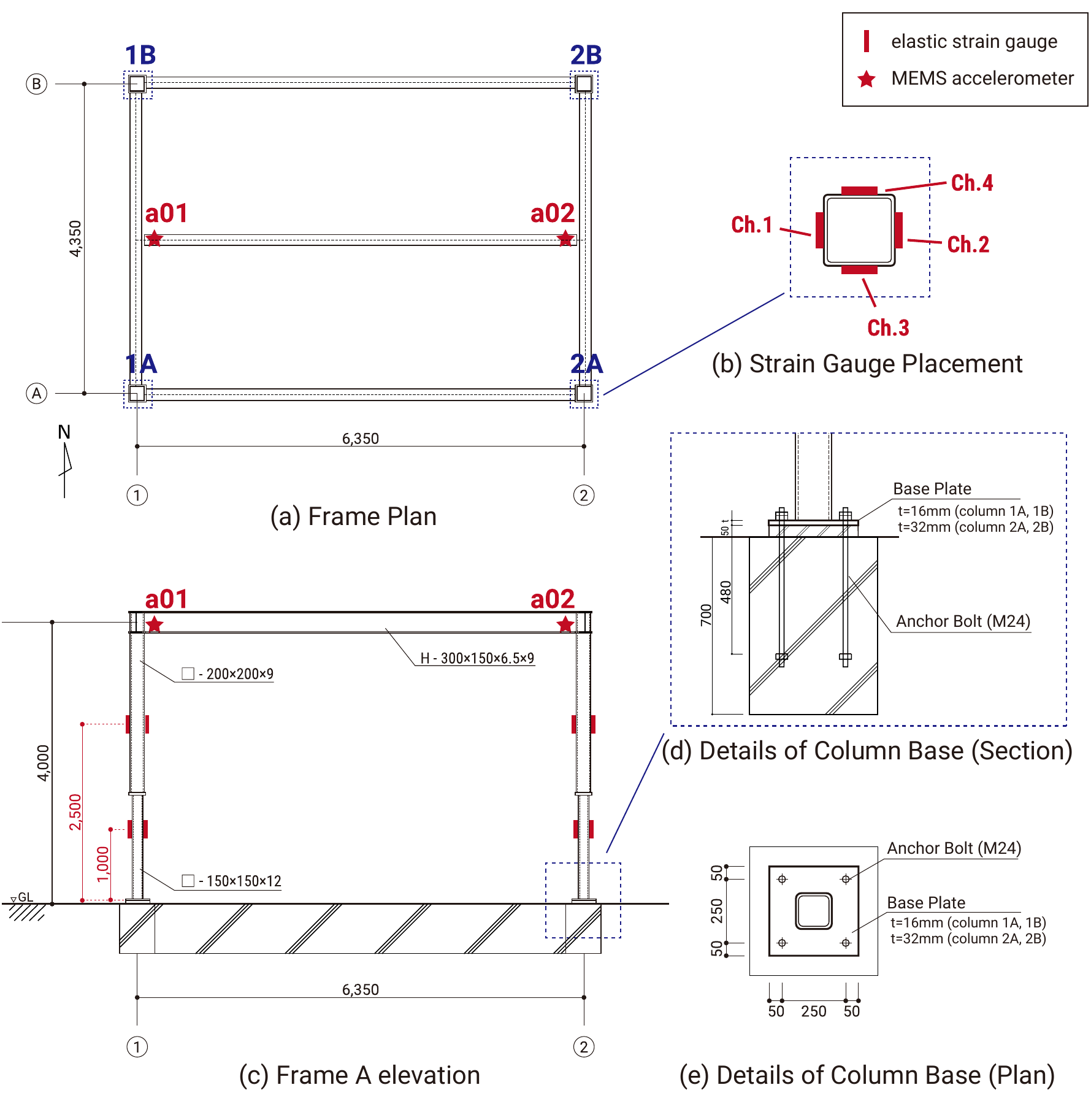}
    \caption{(a) Plan of the experimental model; (b) Details of strain gauge placement; (c) Elevation of Frame A; (d) Details of a column base (Section); (e) Details of a column base (Plan).}\label{fig:view4}
\end{figure}

\section{Target Structure and Data}

\subsection{Full-scale vibration test}

The experimental model, also referred to by Yaoyama et al.\cite{yaoyamaT2023}, is a full-scale, one-story, one-bay by one-bay moment-resisting steel frame, as shown in Figure~\ref{fig:photo}.
The story height is approximately 4,000 mm and the plan is 6,350 mm $\times$ 4,350 mm.
The two frames in the EW direction are called Frames A and B, and those in the NS direction are called Frames 1 and 2.
Accordingly, the four columns are denoted as 1A, 2A, 1B, and 2B, as shown in Figure~\ref{fig:view4}(a).

Each column is composed of two parts with different cross-sectional dimensions. The lower part is 1,500 mm in length and has a box cross section of $\Box-150 \times 150 \times 12$ with $b/t = 12.5$, where $b$ and $t$ represent the width and thickness, respectively.
The upper part is 2,600 mm and has a box cross section of $\Box-200 \times 200 \times 9$ with $b/t = 22.2$.
The material is STKR400, which has a Young's modulus $E = 205,000 ~ \mathrm{N/mm}^2$, yielding strength $245 ~ \mathrm{N/mm}^2$, and maximum strength $400 ~ \mathrm{N/mm}^2$.
The base plate of each column has a different thickness between Frames 1 and 2; the base plates of 1A and 1B have a thickness of 16 mm, and the others have a thickness of 32 mm.
These base plates are connected to the RC footing using four anchor bolts (Figure~\ref{fig:view4}(d) and (e)).

The nominal rotational stiffness $K_\mathrm{BS}$ of these column bases is computed as \cite{aij2021}
\begin{linenomath}\begin{align}
    K_\mathrm{BS} = \frac{E \times n_\mathrm{t} \times A_\mathrm{b} \times (d_\mathrm{t} + d_\mathrm{c})^2}{2l_\mathrm{b}'}
    \label{eq:nominal}
\end{align}\end{linenomath}
where $n_\mathrm{t}$ and $A_\mathrm{b}$ represent the number of anchor bolts on the tensile side and the cross-sectional area of the anchor bolts, respectively. Further, $d_\mathrm{t}$ represents the distance between the centers of the column cross-section and the tensile-side anchor bolts.
In addition, $d_\mathrm{c}$ represents the distance between the center of the column cross-section and the compressive-side edge of the column.
$l_\mathrm{b}' = \min(l_\mathrm{b}, 40d_\mathrm{b})$, where $l_\mathrm{b}$ and $d_\mathrm{b}$ denote the length and diameter of the anchor bolts, respectively.
$E = 205000 ~ \text{N/mm}$; $n_\mathrm{t} = 2$; $A_\mathrm{b} = 452 ~ \text{mm}^2$; $d_\mathrm{t} = 125 ~ \text{mm}$; $d_\mathrm{c} = 75 ~ \text{mm}$; and $l_\mathrm{b}' = 960 ~ \text{mm}$.
The nominal value in this test was computed as $7722 ~ \text{kNm/rad}$.

Considering the interaction effects between a steel beam and slab, the equivalent second moment of the area of the composite beam $I_\mathrm{cs}$ is given by\cite{aij2010}
\begin{linenomath}\begin{align}
    I_\mathrm{cs} = I_\mathrm{s} + (d - x_n)^2 A_\mathrm{s} + \frac{B_\mathrm{e}t}{n} \left(\frac{t^2}{12} + \left(x_n - \frac{t}{2}\right)^2\right)
\end{align}\end{linenomath}
where $B_\mathrm{e}$ denotes the effective width of the slab that contributes to beam rigidification.
$A_\mathrm{s}$ and $I_\mathrm{s}$ represent the area and second moment of the area of the steel beam cross section, respectively;
$t$ represents the effective thickness of the slab; and $x_n$ represents the distance between the upper side of the slab and neutral axis.
Young's modulus ratio $n$ is defined as $E / E_\mathrm{c}$, where $E_\mathrm{c}$ represents the Young's modulus of the concrete. 
According to the recommendations\cite{aij2010}, $B_\mathrm{e}$ is herein given as $B_\mathrm{e} = 585 ~\text{mm}$ for beams in the EW direction or $B_\mathrm{e} = 785 ~\text{mm}$ for those in the NS direction.
With $t = 162.5 ~\text{mm}$, $n = 15$, $I_\mathrm{s} = 7.21 \times 10^7 ~\text{mm}^4$, and $A_\mathrm{s} = 4.68 \times 10^3 ~\text{mm}^2$,
the nominal values of the rigidification ratio were obtained as $EI_\mathrm{cs} / EI_\mathrm{s} = 3.89$ for the beams in the EW-direction and $EI_\mathrm{cs} / EI_\mathrm{s} = 4.28$ for those in the NS-direction.

\begin{table}[!t]
    \centering
    \caption{Vibration tests with different damage cases and excitation directions.}\label{tab:cases}
    \fontsize{9truept}{11truept}\selectfont
    \begin{tabular*}{\textwidth}{@{\extracolsep{\fill}}rlrl}
        \toprule
        Test & Excitation direction & Damage case & Column bases with loosened anchor bolts \\
        \midrule
        1  & EW & 0 & (intact)         \\
        2  & NS & 0 & (intact)         \\
        3  & EW & 1 & 1A               \\
        4  & NS & 1 & 1A               \\
        5  & EW & 2 & 1A, 2A           \\
        6  & NS & 2 & 1A, 2A           \\
        7  & EW & 3 & 1A, 2A, 1B       \\
        8  & NS & 3 & 1A, 2A, 1B       \\
        9  & EW & 4 & 1A, 2A, 1B, 2B   \\
        10 & NS & 4 & 1A, 2A, 1B, 2B   \\
        11 & EW & 5 & all re-tightened \\
        12 & NS & 5 & all re-tightened \\
        \bottomrule
    \end{tabular*}
\end{table}

Two three-axis MEMS accelerometers ADXL355 (Analog Devices), called a01 and a02, were installed on top of the structure, as shown in Figures~\ref{fig:view4}(a) and (c).
Only two lateral components of each accelerometer were used for system identification.
Elastic strain gauges YEFLAB-5 (Tokyo Measuring Instruments Lab.) were attached to all four edges in the box cross-sections of both the lower and upper parts of the columns, as shown in Figures~\ref{fig:view4}(b) and (c).
The four strain components at the cross-section are called Ch.~1--4 as shown in Figure~\ref{fig:view4}(c).
Four acceleration measurement components (two components for each accelerometer) and 32 strain measurement components (four components for each cross section and two cross-sections for each column) were used in this study. These signals were collected using Raspberry Pi units connected to WiFi networks.

The manual excitation of the structure yielded free-vibration responses and considered six damage cases (Cases 0--5). Case 0 represented the initial intact state. Case 1 represented a damaged state in which the anchor bolts connecting column 1A to its base plate were loosened to simulate a decrease in stiffness.
Cases 2--4 represented gradually worsening damage states, in which the anchor bolts of column bases 2A, 1B, and 2B were loosened subsequently.
Case 5 represented a re-simulated intact state in which the anchor bolts of all column bases were re-tightened.
Two tests in different excitation directions (the EW- and NS- directions) were conducted for each of the aforementioned damage cases, resulting in a total of 12 tests (Tests 1--12), as listed in Table 1.

\subsection{Pre-processing data}

\begin{figure}[!t]
    \centering
    \includegraphics[width=150truemm]{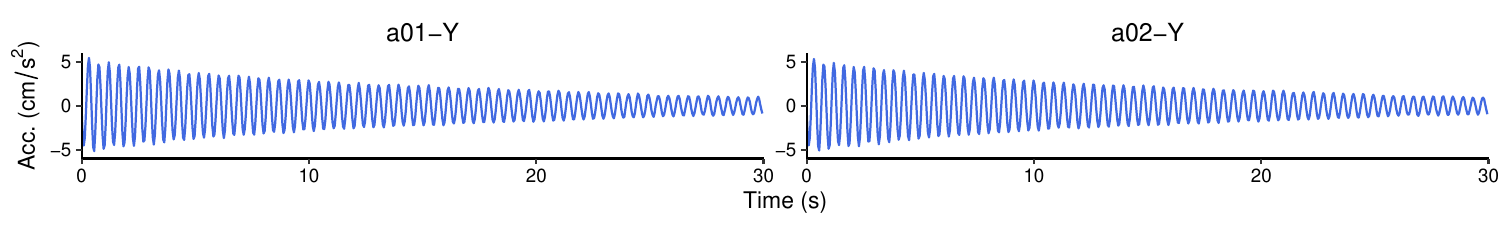}
    \caption{Time histories of measured acceleration in Test 1 (Case 0, EW-direction excitation).}\label{fig:hist_acc}
\end{figure}

\begin{figure}[!t]
    \centering
    \includegraphics[width=150truemm]{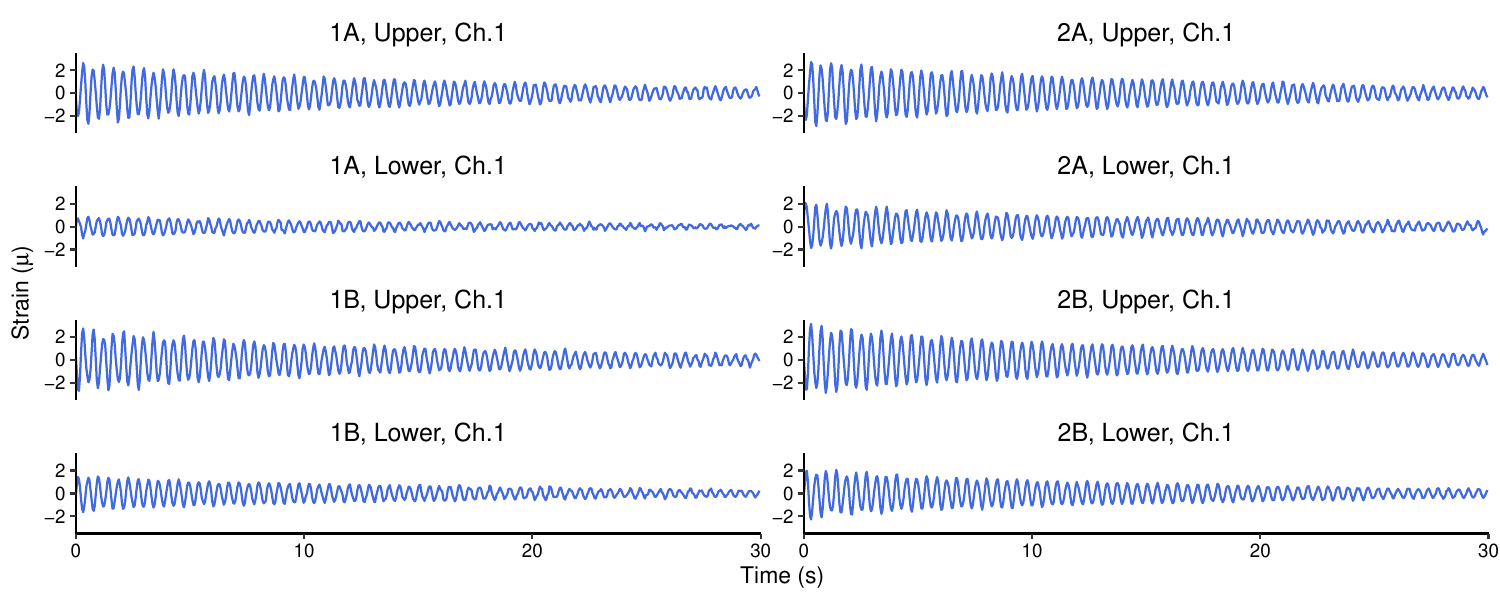}
    \caption{Time histories of measured strain for Ch. 1 in Test 1 (Case 0, EW-direction excitation).}\label{fig:hist_str}
\end{figure}

Time synchronization of the measured signals was performed based on linear interpolation using the timestamps of each unit. This was first performed individually for the strain gauges and accelerometers because of their different sampling rates (87 and 125 Hz, respectively).
A cosine-tapered bandpass filter was used to reduce noise in the low- ($<$ 1 Hz) and high-frequency ($>$ 10 Hz) bands.
Downsampling based on linear interpolation was performed such that all signals, including the strain and acceleration responses, were synchronized and had the same sampling period, $\Delta t = 0.046 ~(\text{s})$.
For Test 1 (Case 1, EW-direction excitation), Figures~\ref{fig:hist_acc} and \ref{fig:hist_str} show the pre-processed time histories
for acceleration in the EW-direction (Figure~\ref{fig:hist_acc}), and the strain at Ch.~1 (Figure~\ref{fig:hist_str}).

\section{Proposed Framework}

\subsection{Output-only system identification}

This study considered the free-vibration responses of the test structure to manual excitation, and therefore, a method for output-only system identification, i.e., subspace state-space system identification (4SID), is adopted for modal parameter estimation.
This system identification approach considers an output-only linear time-invariant (LTI) system described by 
\begin{linenomath}\begin{align}
    \mathbf{x}(t+1) &= \mathbf{A}\mathbf{x}(t) + \mathbf{w}(t) \label{eq:system} \\
    \mathbf{y}(t) &= \mathbf{C}\mathbf{x}(t) + \mathbf{u}(t) \label{eq:observation}
\end{align}\end{linenomath}
where $\mathbf{x} \in \mathbb{R}^n$, $\mathbf{y} \in \mathbb{R}^p$, and  $\mathbf{w}(t) \in \mathbb{R}^n$ represent the state vector, output vector (measured acceleration and strain), and system noise vector, respectively. Further,  $\mathbf{v}(t) \in \mathbb{R}^p$ represents the observation noise vector. $\mathbf{A} \in \mathbb{R}^{n \times n}$ and $\mathbf{C} \in \mathbb{R}^{p \times n}$ are constants, and hereafter, they are referred to as \textit{system matrices}.
4SID is used to obtain system matrices, and to this end, the stochastic realization theory was employed (see Lemma 7.9 in the work by Katayama\cite{katayamaT2005} for details).

Using $\mathbf{A}$ and $\mathbf{C}$, the modal parameters of the natural frequencies, damping ratios, and mode shapes are expressed by 
\begin{linenomath}\begin{align}
    \omega_j &= \frac{|\log \lambda_j|}{\Delta t} \\
    \zeta_j &= - \frac{\log |\lambda_j|}{|\log \lambda_j|} \\
    \mathbf{u}_j &= \mathbf{C} \mathbf{v}_j
\end{align}\end{linenomath}
where $\lambda_j$ and $\mathbf{v}_j \in \mathbb{R}^n$ represent the $j$-th eigenvalue and eigenvector of $\mathbf{A}$, and $\Delta t$ represents the sampling period.

Critical hyperparameters that should be carefully adjusted in the implementation include the number of system orders $n$ (i.e., the length of the system vector $\mathbf{x}$), the number of block rows of the block Hankel matrix\cite{katayamaT2005} $k$, and the length of the time series extracted for identification. $n$ is a critical parameter because it is related to the number of identified vibration modes.
A successful 4SID provides conjugate pairs of complex eigenvectors, and thus, in a practical sense, results in the $n/2$ vibration modes.

When responses to seismic ground motions are obtained, the same formulation can be enabled by replacing Eqs.~(\ref{eq:system}) and (\ref{eq:observation}) with a description of a multi-input multi-output (MIMO) system.
System matrices can be obtained using MIMO subspace identification methods such as MOESP\cite{verhaegenM1992a,verhaegenM1992b} and N4SID\cite{vanoverscheeP1994}.

\subsection{Modal displacements and bending moments}

The identified mode shapes in terms of strain and acceleration responses can be used to express the displacement--stress relationship in the modal space\cite{iyamaJ2021,iyamaJ2023,yaoyamaT2023}.

First, the relationship between the identified acceleration mode shape called \textit{modal acceleration}, $a$, and the corresponding displacement mode shape called \textit{modal displacement} (MD), $d$ is given by
\begin{linenomath}\begin{align}
    d = - a / \omega^2
\end{align}\end{linenomath}
where $\omega$ represents the corresponding natural angular frequency obtained through system identification.

Second, using the \textit{modal strains} at a cross-section, the bending-moment mode shape called the \textit{modal bending moment} (MBM), denoted by $m$, can be computed as 
\begin{linenomath}\begin{align}
    m = (\varepsilon_\mathrm{lo} - \varepsilon_\mathrm{up}) E Z / 2
\end{align}\end{linenomath}
where $\varepsilon_\mathrm{lo}$ and $\varepsilon_\mathrm{up}$ represent the modal strains at the upper and lower sides of the cross-section, respectively, and $E$ and $Z$ represent the Young's modulus and section modulus, respectively.
All aforementioned mode-shape values are complex in the following examination. However, after confirming that the modal strains of the upper and lower sides have almost opposite phases, and that the modal strains and corresponding modal accelerations have almost the same or opposite phases,
their absolute values, provided with appropriate signs, are used as $\varepsilon_\mathrm{lo}, \varepsilon_\mathrm{up}$ and $d$.

Finally, the normalized MBM is introduced as 
\begin{linenomath}\begin{align}
    \tilde{m} = m / d_\mathrm{ref}
\end{align}\end{linenomath}
where $d_\mathrm{ref}$ denotes the reference modal displacement (MD) value.
In this study, the MD value for the lateral displacement of the midpoint of the beam was set as $d_\mathrm{ref}$.
For Frames A and B, the mean value of the MDs for the EW-direction components of a01 and a02 was used as $d_\mathrm{ref}$.
For Frames 1 and 2, the MDs for the NS-direction component of a01 and a02 were used, respectively.
Thus, the normalized MBMs denote bending moments when a lateral force is applied such that the midpoint of the beam has a unit lateral displacement.
These values represent the displacement--stress relationship, which directly provides the knowledge of stiffness without any mass assumption.

\subsection{Hierarchical Bayesian approach for multitask learning}

This subsection provides a general description of the proposed multitask learning framework based on HBM.
Although the following formulation focuses on strain and acceleration measurements for a steel frame structure, it can easily be extended to various types of structures, e.g., by employing substructure approaches\cite{papadimitriouC2013}. Such generalization will be studied in the future.

Consider a space steel frame composed of $F$ planar frames with strain and acceleration measurements for each frame.
For the $f$th frame, let ${}_f\bm{\theta}$ and ${}_f\tilde{\bm{m}}$ represent a parameter vector composed of the stiffness of columns or beams of the frame and a vector of normalized MBMs identified from the measurements, respectively.
The combined vectors are defined as $\bm{\theta} = \{{}_1\bm{\theta}^\top,...,{}_F\bm{\theta}^\top\}^\top$ and $\tilde{\bm{m}} = \{{}_1\tilde{\bm{m}}^\top,...,{}_F\tilde{\bm{m}}^\top\}^\top$.
The probability that data $\tilde{\bm{m}}$ are observed under a structural model represented by $\bm{\theta}$, i.e., the likelihood, is assumed to be
\begin{linenomath}\begin{align}\label{eq:likelihood}
    p(\tilde{\bm{m}} \mid \bm{\theta}, \bm{\beta}) &= \prod_{f=1}^F p({}_f\tilde{\bm{m}} \mid {}_f\bm{\theta}, {}_f\beta) \nonumber \\
    &= \prod_{f=1}^F \mathcal{N}({}_f\tilde{\bm{m}} \mid {}_f\hat{\bm{m}}({}_f\bm{\theta}), {}_f\beta^{-1} \mathbf{I})
\end{align}\end{linenomath}
where $\bm{\beta} = \{{}_1\beta, ..., {}_F\beta\}^\top$ represents a vector of precision parameters shared across measurements at different sections within a frame.
$\hat{\bm{m}}({}_f\bm{\theta})$ represents a simulated normalized MBM given parameter ${}_f\bm{\theta}$.
$\mathbf{I} \in \mathbb{R}^{F \times F}$ denotes the identity matrix, and 
$\mathcal{N}(. \mid \bm{\mu}, \mathbf{\Sigma})$ denotes a multivariate Gaussian distribution with mean $\bm{\mu}$ and covariance $\mathbf{\Sigma}$.
Eq.~(\ref{eq:likelihood}) assumes that the prediction error for the $f$th frame, denoted by ${}_f\bm{e} = {}_f\tilde{\bm{m}} - {}_f\hat{\bm{m}}$, independently follows a zero-mean multivariate Gaussian distribution.

Consider the case wherein some elements of parameter vector ${}_f \bm{\theta}$ are assumed to have similarities between different frames.
Let ${}_f \bm{\theta}^\ast$ denote a vector of such parameters, and ${}_f \bm{\theta}^{-\ast}$ represent a vector of the other parameters.
Then, the combined vectors $\bm{\theta}^\ast$ and $\bm{\theta}^{-\ast}$ are defined.
Utilizing the knowledge of similarities across different \textit{tasks}, that is, different frames, multitask learning is formulated and tackled via HBM.
A hierarchical prior over $\bm{\theta}^\ast$ is introduced as
\begin{linenomath}\begin{align}\label{eq:hierarchicalprior}
    p(\bm{\theta}^\ast \mid {}_0\bm{\theta}^\ast, \bm{\alpha}) &= \prod_{f=1}^F p({}_f\bm{\theta}^\ast \mid {}_0\bm{\theta}^\ast, \bm{\alpha}) \nonumber \\
    &= \prod_{f=1}^F \mathcal{N}({}_f\bm{\theta}^\ast \mid {}_0\bm{\theta}^\ast, \mathrm{diag}(\bm{\alpha})^{-1})
\end{align}\end{linenomath}
where ${}_0\bm{\theta}^\ast, \bm{\alpha}$ are hyperparameters, and $\mathrm{diag}(\cdot)$ represents a diagonal matrix.
Eq.~(\ref{eq:hierarchicalprior}) indicates that the parameters ${}_f\bm{\theta}^\ast$ of different frames are generated individually from a common prior, which encourages different frames to transfer the knowledge of the parameters to each other.

Based on the aforedescribed model, the posterior probability of all parameters is obtained by  
\begin{linenomath}\begin{align}\label{eq:fullposeterior}
    ~ & p(\bm{\theta}, {}_0\bm{\theta}^\ast, \bm{\beta}, \bm{\alpha} \mid \tilde{\bm{m}}) \nonumber \\
    \propto ~ & p(\tilde{\bm{m}} \mid \bm{\theta}, \bm{\beta}) ~ p(\bm{\theta}^\ast \mid {}_0\bm{\theta}^\ast, \bm{\alpha}) ~ p({}_0\bm{\theta}^\ast) ~ p(\bm{\theta}^{-\ast}) ~ p(\bm{\beta}) ~ p(\bm{\alpha}) \nonumber \\
    = ~ & \prod_{f=1}^F \mathcal{N}({}_f\tilde{\bm{m}} \mid {}_f\hat{\bm{m}}({}_f\bm{\theta}), {}_f\beta^{-1} \mathbf{I})
    ~ \mathcal{N}({}_f\bm{\theta}^\ast \mid {}_0\bm{\theta}^\ast, \mathrm{diag}(\bm{\alpha})^{-1}) ~ p({}_0\bm{\theta}^\ast) ~ p(\bm{\theta}^{-\ast}) ~ p(\bm{\beta}) ~ p(\bm{\alpha})
\end{align}\end{linenomath}
To evaluate the aforedescribed posterior, a Markov chain Monte Carlo (MCMC) sampling was adopted in this study, as in \cite{beckJL2002,chingJ2007,rocchettaR2018,huangY2017b}.

\subsection{Application to a one-story, one-bay by one-bay steel frame}

\begin{figure}[!t]
    \centering
    \includegraphics[width=80truemm]{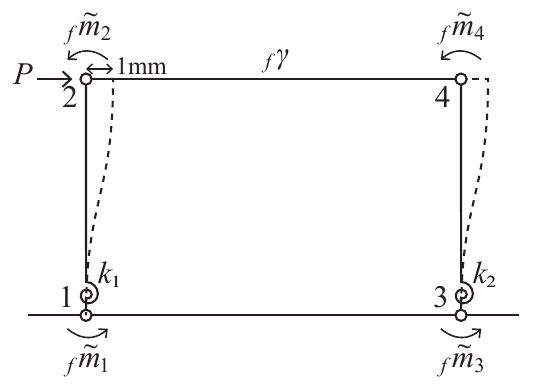}
    \caption{Illustration of the structural model assumed for each planar frame.}\label{fig:updated_model}
\end{figure}

When applying the abovementioned framework to a one-story, one-bay by one-bay steel frame, a moment-resisting frame with two rotational springs used to represent semi-rigid column bases, as shown in Figure~\ref{fig:updated_model},
is assumed for each frame $f \in \{\text{A, B, 1, 2}\}$ and updated to minimize the discrepancy between the identified and simulated normalized MBM values.
The model parameter vector for each frame, denoted by ${}_f\bm{\theta}$, is composed of rotational spring stiffness values ${}_fk_i~(i = 1, 2)$ and the rigidification coefficient ${}_f\gamma$.
This coefficient scales the nominal bending stiffness of the steel beam ${}_f EI_\mathrm{s}$ to the actual value ${}_f {EI}_\mathrm{cs}$
considering the beam-slab interaction as ${}_f {EI}_\mathrm{cs} = 10^{{}_f\gamma} \times {}_f EI_\mathrm{s}$.
For simplicity, the mechanical and material properties of the other elements are fixed at their nominal values.
In this application, the vector $\tilde{\bm{m}}$ is composed of ${}_f\tilde{\bm{m}} ~ (f \in \{\mathrm{A},\mathrm{B},\mathrm{1},\mathrm{2}\})$, where each of them is composed of normalized MBM values at four different sections, ${}_f\tilde{m}_i ~ (i = 1,...,4)$, as shown in Figure~\ref{fig:updated_model}.

According to Eq.~(\ref{eq:likelihood}), ${}_f\tilde{m}_i$ at different sections individually follow the likelihood
\begin{linenomath}\begin{align}
    {}_f\tilde{m}_i \mid {}_f\bm{\theta}, {}_f\sigma_m \sim \mathcal{N}({}_f\hat{m}_i({}_fk_1, {}_fk_2, {}_f\gamma), {}_f\sigma_m^2)
\end{align}\end{linenomath}
where ``$a \mid b \sim p$'' means ``$a$ conditioned on $b$ is sampled from $p$.''
The standard deviation (s.d.) of the observation noise ${}_f\sigma_m$ involved in the measured normalized MBMs is shared in the same vibration test; that is,
${}_\mathrm{A}\sigma_m = {}_\mathrm{B}\sigma_m \equiv {}_\mathrm{AB}\sigma_m$ (for the EW-direction tests) and ${}_1\sigma_m = {}_2\sigma_m \equiv {}_{12}\sigma_m$ (for the NS-direction tests).

Assuming that the rigidification coefficients $\{{}_f\gamma\}$ are similar for different frames, hierarchical priors in Eq. (\ref{eq:hierarchicalprior}) can be expressed as
\begin{linenomath}\begin{align}
    {}_f\gamma \mid {}_0\gamma, \sigma_\gamma \sim \mathcal{N} ({}_0\gamma, \sigma_\gamma^2) \label{eq:hierarchical}
\end{align}\end{linenomath}
The rigidification coefficients of all frames are generated from a common normal prior that maintains values proximal to the prior mean ${}_0\gamma$, 
where the degree of proximity is adjusted using prior variance $\sigma_\gamma^2$.
In contrast, an uninformative prior is given for ${}_fk_i$ rather than a hierarchical prior, assuming that the stiffness of the column bases varies individually because of structural damage.

To ensure sampling convergence, \textit{weakly informative priors}\cite{gelmanA2008} are set over the hyperparameters ${}_f\sigma_m, {}_0\gamma, \sigma_\gamma$ as
\begin{linenomath}\begin{align}
    {}_\mathrm{AB}\sigma_m &\sim \mathcal{N}^+ (0, 100^2) \label{eq:f_sigma_mAB} \\
    {}_{12}\sigma_m &\sim \mathcal{N}^+ (0, 100^2) \label{eq:f_sigma_m12} \\
    {}_0\gamma &\sim \mathcal{N}^+ (0, 3^2) \label{eq:0_gamma} \\
    \sigma_\gamma &\sim \mathcal{N}^+ (0, 5^2) \label{eq:sigma_gamma}
\end{align}\end{linenomath}
where $\mathcal{N}^+(.,.)$ denotes a half-normal distribution; that is, a truncated normal distribution defined over a random variable greater than or equal to zero.
For example, the weakly informative prior in Eq. (\ref{eq:f_sigma_m12}) assumes that the s.d. of the observation noise for Frames 1 and 2 is not likely to exceed 100 (Nm/mm), although it allows exceedance with a small probability.
Thus, hierarchical modeling helps us to adopt a priori knowledge in a flexible manner while preventing strong bias.

\begin{figure}[!t]
    \centering
    \includegraphics[width=100truemm]{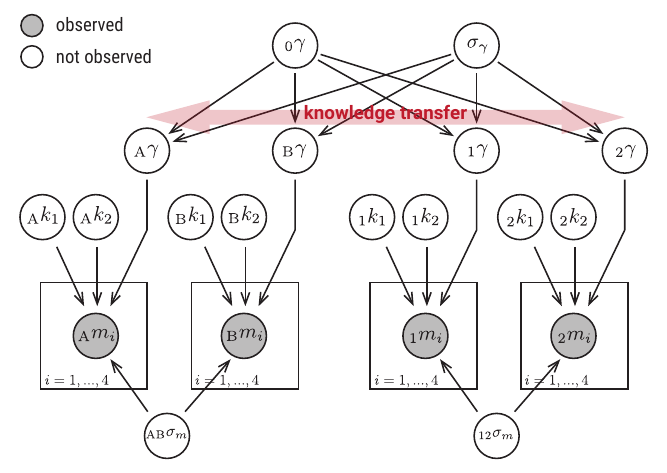}
    \caption{Graphical model representing the hierarchical model assumed in this study.}\label{fig:graphical_models}
\end{figure}

Figure~\ref{fig:graphical_models} illustrates the graphical model used to represent the hierarchical modeling assumed in this study. The gray and white vertices represent the observed and unobserved random variables, respectively, and the edges between the vertices represent their dependencies.
Each box, called a plate, is a reduced expression representing multiple vertices of the same kind; for example, the box corresponding to ${}_\mathrm{A} m_i$ represents a set of variables ${}_\mathrm{A} m_1, ..., {}_\mathrm{A} m_4$.

Hyperparameters ${}_0\gamma, \sigma_\gamma$ are sequentially updated from Cases 0 to 5,
assuming that measurements are subsequently obtained and all past measurements are used for each case (\textit{sequential, multitask (SMT) inference scenario}).
In this scenario, the mean and s.d. of the posterior samples of ${}_0\gamma, \sigma_\gamma$ in Case $i$ are used as those of the prior half-normal distribution in Case $i + 1$.
This promotes knowledge transfer across different frames and datasets (i.e., damage cases).
Such knowledge transfer is neither adopted for the rotational stiffness parameters ${}_fk_1, {}_fk_2$ because informative priors on these parameters decrease the sensitivity to structural damage, nor is it used for the observation noise parameters ${}_f\sigma_\mathrm{AB}, {}_f\sigma_\mathrm{12}$, which are strongly affected by variations in the experimental conditions depending on the tests.
Another inference scenario is considered for comparison, where only a single set of measurements is obtained for each case; that is, the parameters are independently updated, and no knowledge transfer regarding ${}_f\gamma$ between frames is assumed (\textit{Individual, Single-Task (IST) inference scenario}).
This scenario does not use the hierarchical prior expressed in Eq.~(\ref{eq:hierarchical}); instead, it adopts weakly informative priors ${}_f\gamma \sim \mathcal{N}^+(0, 5^2)$ assumed for all frames and cases.

Fully Bayesian inference was performed using an MCMC algorithm, i.e., the NUTS sampler\cite{hoffmanMD2014}, which is implemented in the R package \textit{cmdstanr}\cite{cmdstanr}, as an interface to \textit{Stan}\cite{stan}.
Four Markov chains with different initial values and 10,000 samples were generated, and the burn-in period of 5,000 samples for each chain was discarded; 20,000 samples were obtained as the posterior samples.
Convergence was judged for all parameters using a convergence metric $\hat{R}$, where the threshold was set as $\hat{R} < 1.1$ according to Gelman et al.\cite{gelmanA2013}.
The MCMC simulation was performed with a PC having an Intel Core i7-1065G7 CPU at 1.30 GHz (using four cores) and 16 GB of RAM. The average runtime for Cases 0--5 was 243 s for the SMT scenario and 295 s for the IST scenario.

\section{Results}

\subsection{Modal identification}

\begin{figure}[!t]
    \centering
    \includegraphics[width=150truemm]{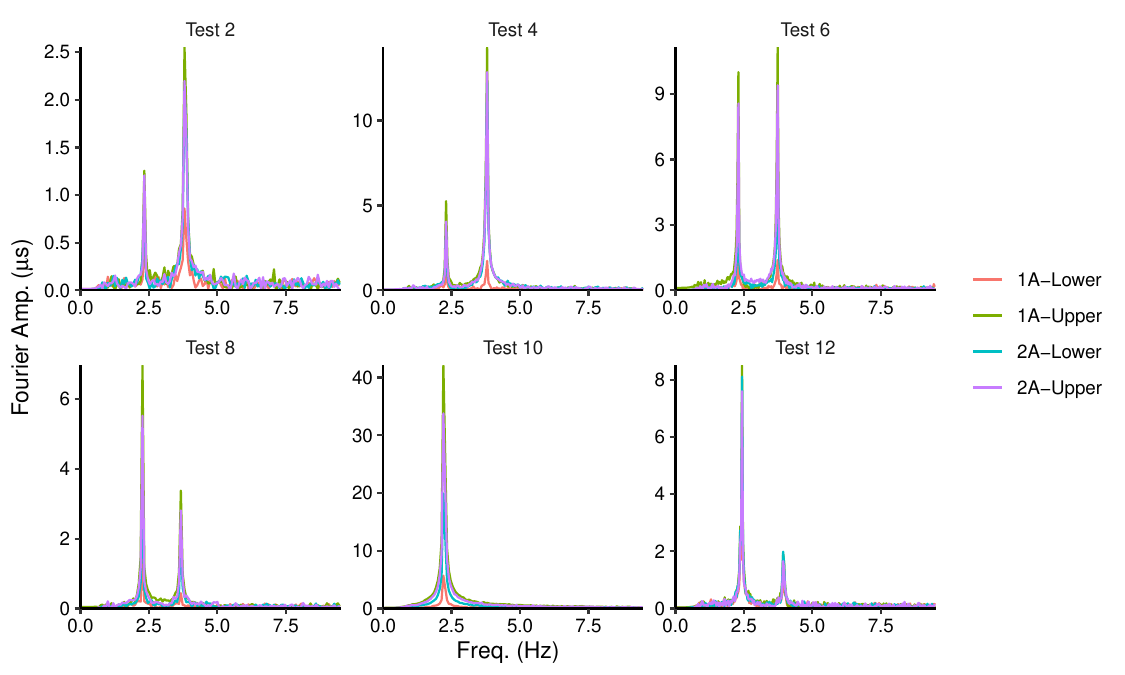}
    \caption{Fourier amplitudes of strain responses in the NS-direction excitation tests.}\label{fig:fourier}
\end{figure}

System identification based on stochastic realization theory\cite{katayamaT2005} was performed for each test using the strain and acceleration responses.
The number of system orders was determined to be $n = 6$ by preliminary examinations, which means that three vibration modes (Modes 1--3) were obtained.
However, for Test 10 (Case 4, NS-direction excitation), $n = 10$ was used because a preliminary examination adopting $n = 6$ (and $n = 8$) failed to fully identify the modes that appear to correspond to Modes 1--3.
To examine this, Figure~\ref{fig:fourier} shows the Fourier amplitudes of the strain responses measured at Ch.~3 for both the upper and lower cross-sections of columns 1A and 2A, for the six tests in the NS direction.
Test 10 has a peak only around 2.5 Hz, unlike the other tests that have clear peaks both around 2.5 and 4.0 Hz.
This can be attributed to the insufficient excitation of the mode around 4.0 Hz in the test. The first, third, and fourth modes identified for Test 10 with $n = 10$, which have the closest values to the identified natural frequencies in Test 7 (Case 4, EW-direction excitation), were used and defined as Modes 1, 2, and 3, respectively.

The number of block rows in the block Hankel matrix was set to $k = 30$ for all tests. The time series used for the identification were carefully extracted for each test.

\begin{figure}[!t]
    \centering
    \includegraphics[width=150truemm]{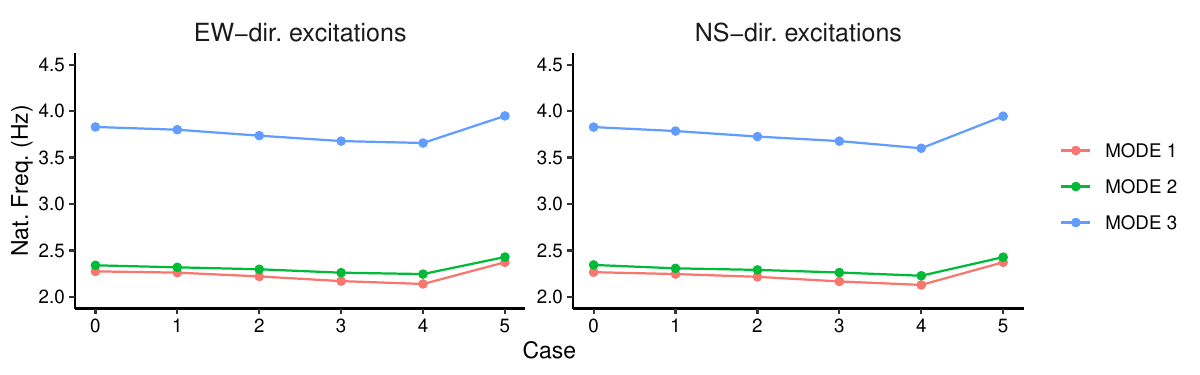}
    \caption{Identified natural frequencies for all cases and both excitation directions.}\label{fig:freq}
\end{figure}

\begin{figure}[!t]
    \centering
    \includegraphics[width=150truemm]{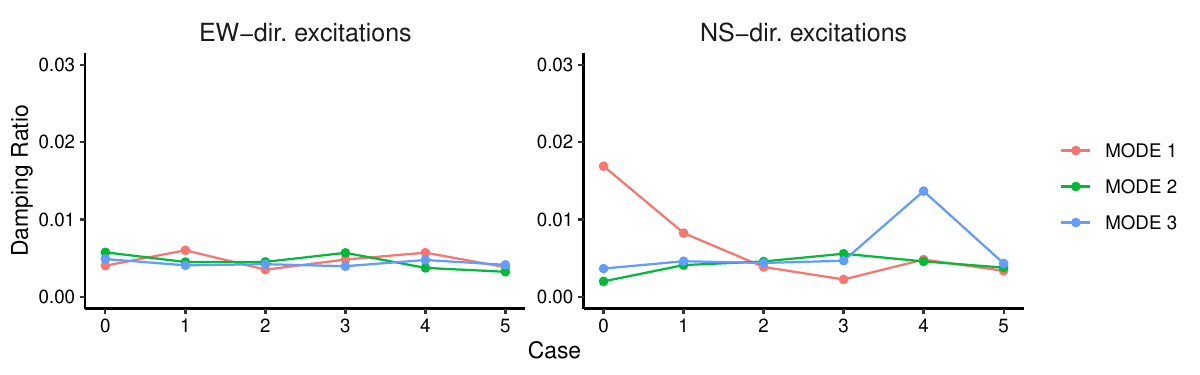}
    \caption{Identified damping ratios for all cases and both excitation directions.}\label{fig:damp}
\end{figure}

Figure~\ref{fig:freq} shows the identified natural frequencies of Modes 1--3 for all cases (Cases 0--5) and in both directions (EW and NS).
Two modes (Modes 1 and 2) around 2.3 Hz and one mode (Mode 3) around 3.8 Hz were found in both directions. The successful identification of the two proximal modes (Modes 1 and 2), which may not be visually distinguished from the Fourier amplitude spectra, demonstrates the high performance of the identification algorithm. In both directions, the natural frequencies of all modes decreased from Cases 0 to 4 and increased from Cases 4 to 5, which corresponds to gradually worsening damage states and a resimulated intact state.
Figure~\ref{fig:damp} presents the identified damping ratios.
In the EW-direction excitation tests, all modes had similar damping ratios of approximately 0.005. In the NS-direction excitation tests, Modes 1 and 3 had relatively unstable damping ratios, whereas Mode 2 exhibits more stable values. This mode corresponds to the NS-direction translational mode and is considered significantly excited in the NS-direction tests.

\begin{figure}[!t]
    \centering
    \includegraphics[width=150truemm]{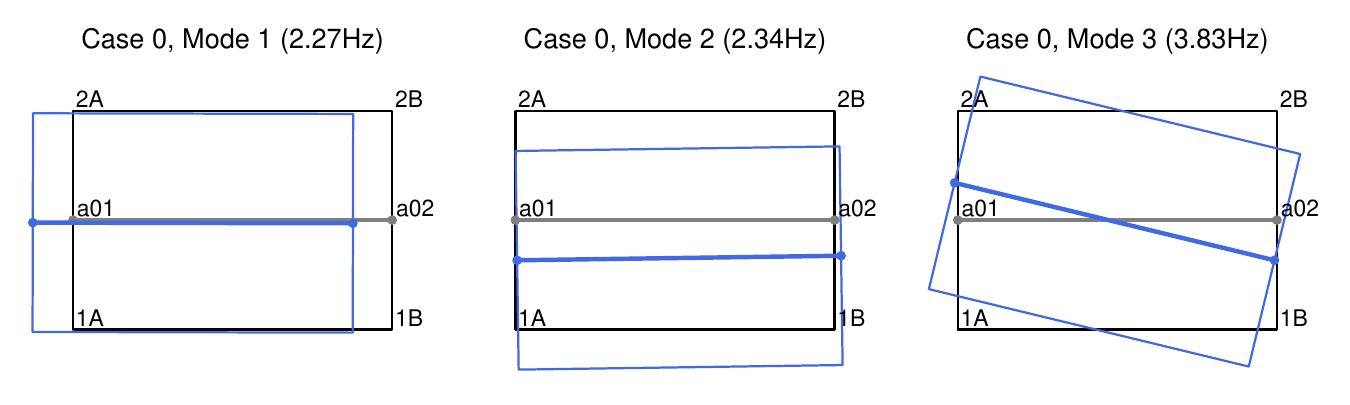}
    \caption{Identified displacement mode shapes (only real part) for Modes 1--3 in Test 1.}\label{fig:modes}
\end{figure}

Figure~\ref{fig:modes} shows a plan view of the real parts of the acceleration (displacement) mode shapes identified in Test 1.
This indicates that Modes 1 and 2 represent the translational modes in the EW and NS directions, respectively, and Mode 3 represents the planar rotational mode.

\begin{figure}[!t]
    \centering
    \includegraphics[width=150truemm]{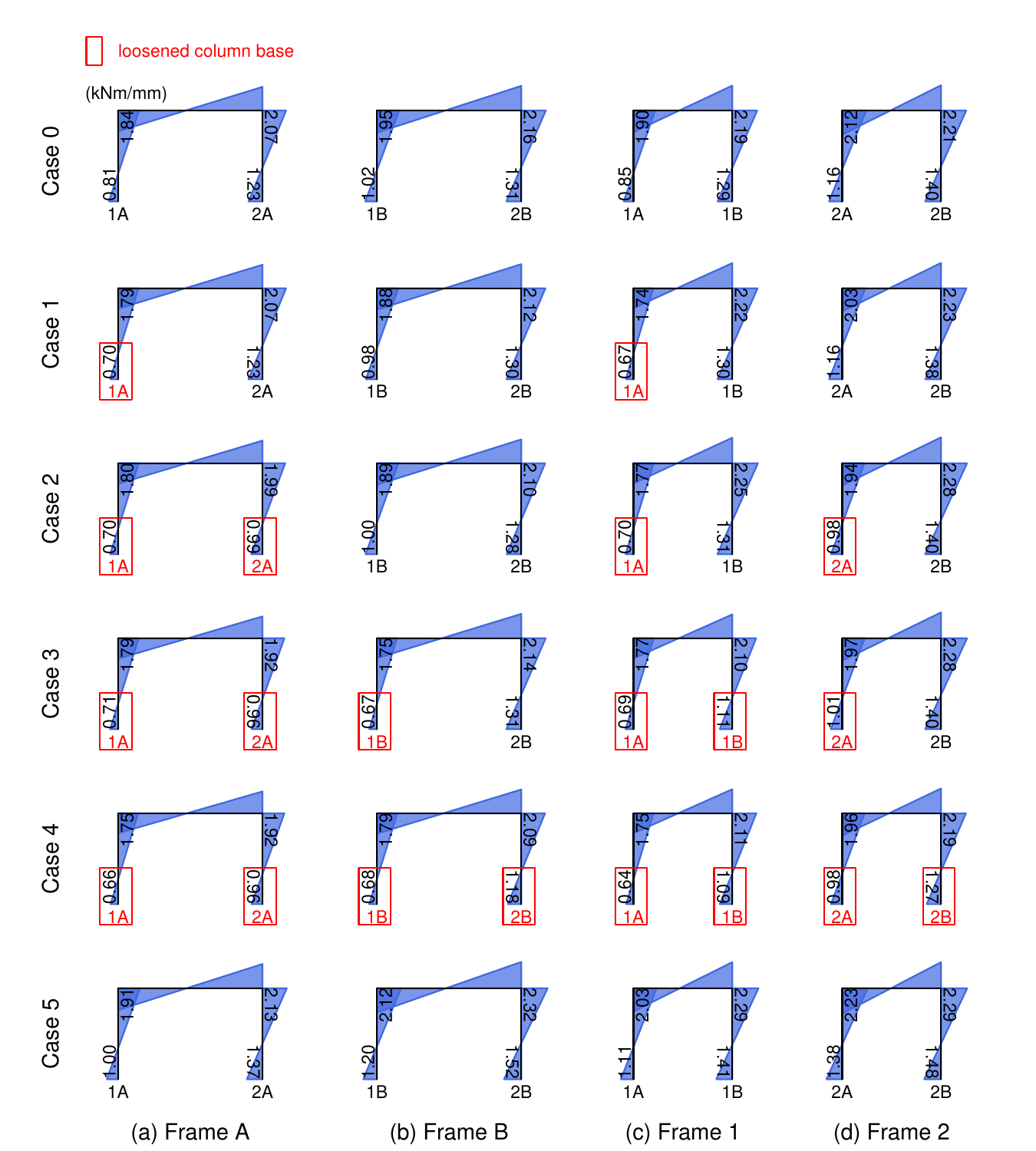}
    \caption{Identified normalized modal bending moment diagrams for all frames in all cases.}\label{fig:mbms}
\end{figure}

Figure~\ref{fig:mbms} shows the normalized MBMs (Nm/mm) identified for all frames in all cases. The MBMs of the EW- and NS-direction frames were computed using Modes 1 and 2 in the EW- and NS-direction tests, respectively.
The decrease in the stiffness of the column bases corresponds to the decrease in MBMs at the bases in all frames. From Case 0 to 1, loosening the anchor bolts of the base of column 1A reduces the MBM at the base by more than 10\% (0.81 $\rightarrow$ 0.70). However, it has almost no effect on the MBMs at the other column bases, which implies that the sensitivity and robustness of MBMs are expected to allow an effective model updating of steel frame structures.

\subsection{Model updating results}

\begin{figure}[!t]
    \centering
    \includegraphics[width=120truemm]{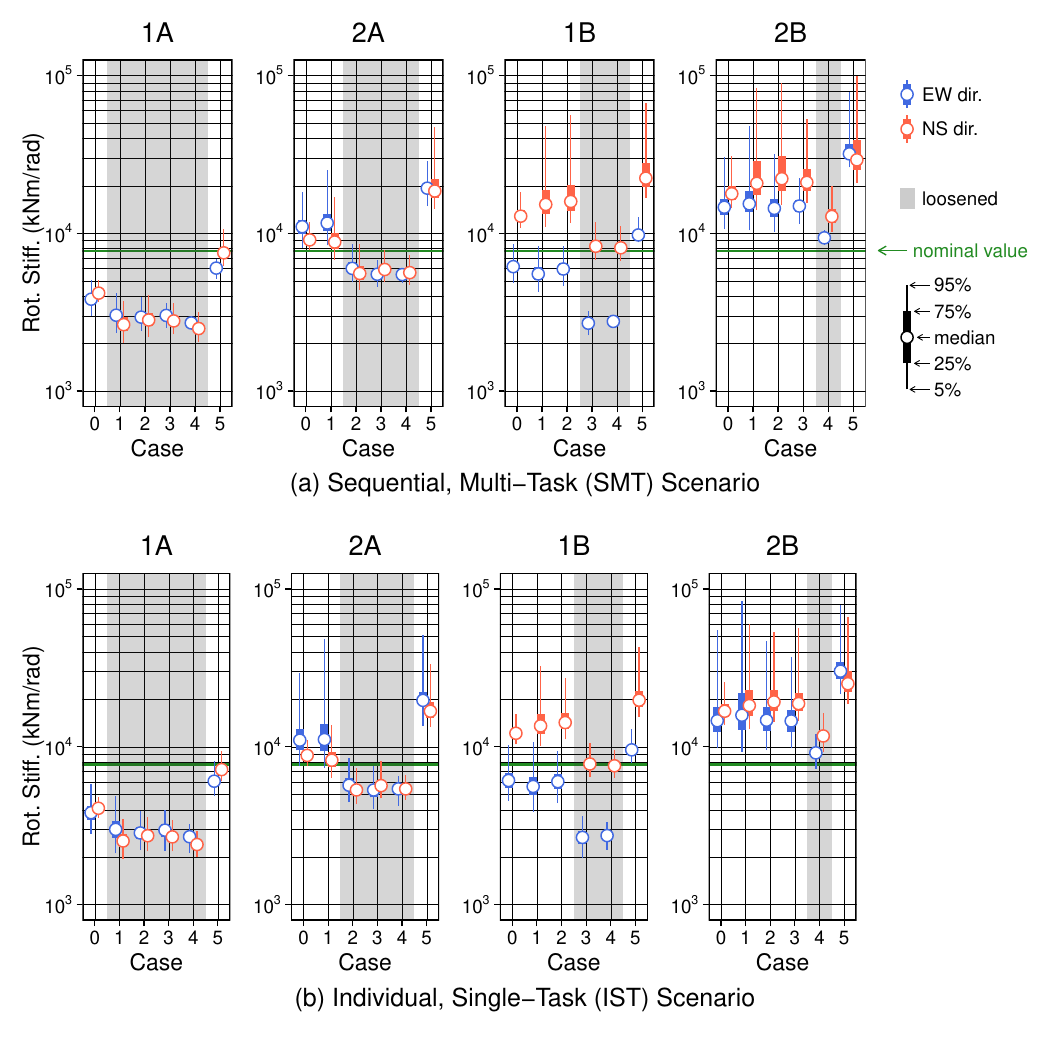}
    \caption{Posterior distributions of rotational stiffness (${}_fk_1$ and ${}_fk_2$) of all column bases in both directions for Cases 0--5, for (a) sequential multitask inference scenario and (b) individual, single-task inference scenario.}\label{fig:kbs}
\end{figure}

\begin{figure}[!t]
    \centering
    \includegraphics[width=160truemm]{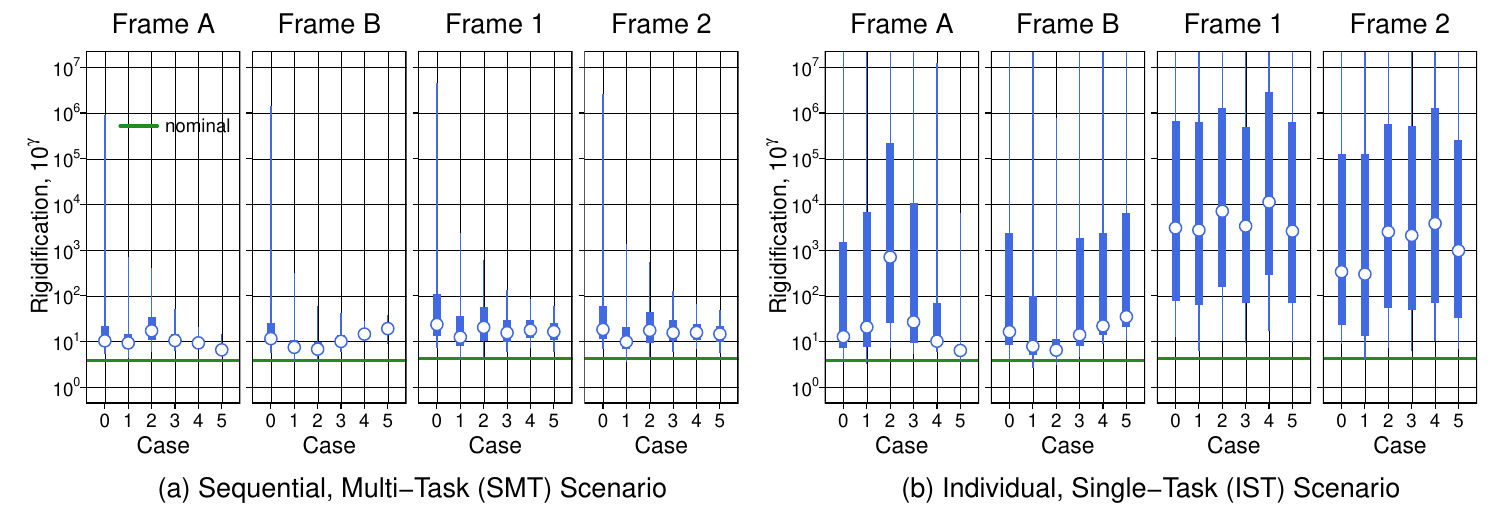}
    \caption{Posterior distributions of beam rigidification coefficients transformed as $10^{{}_f\gamma}$ for (a) sequential, multitask inference scenario and (b) individual, single-task inference scenario.}\label{fig:bmf}
\end{figure}

\begin{figure}[!t]
    \centering
    \includegraphics[width=160truemm]{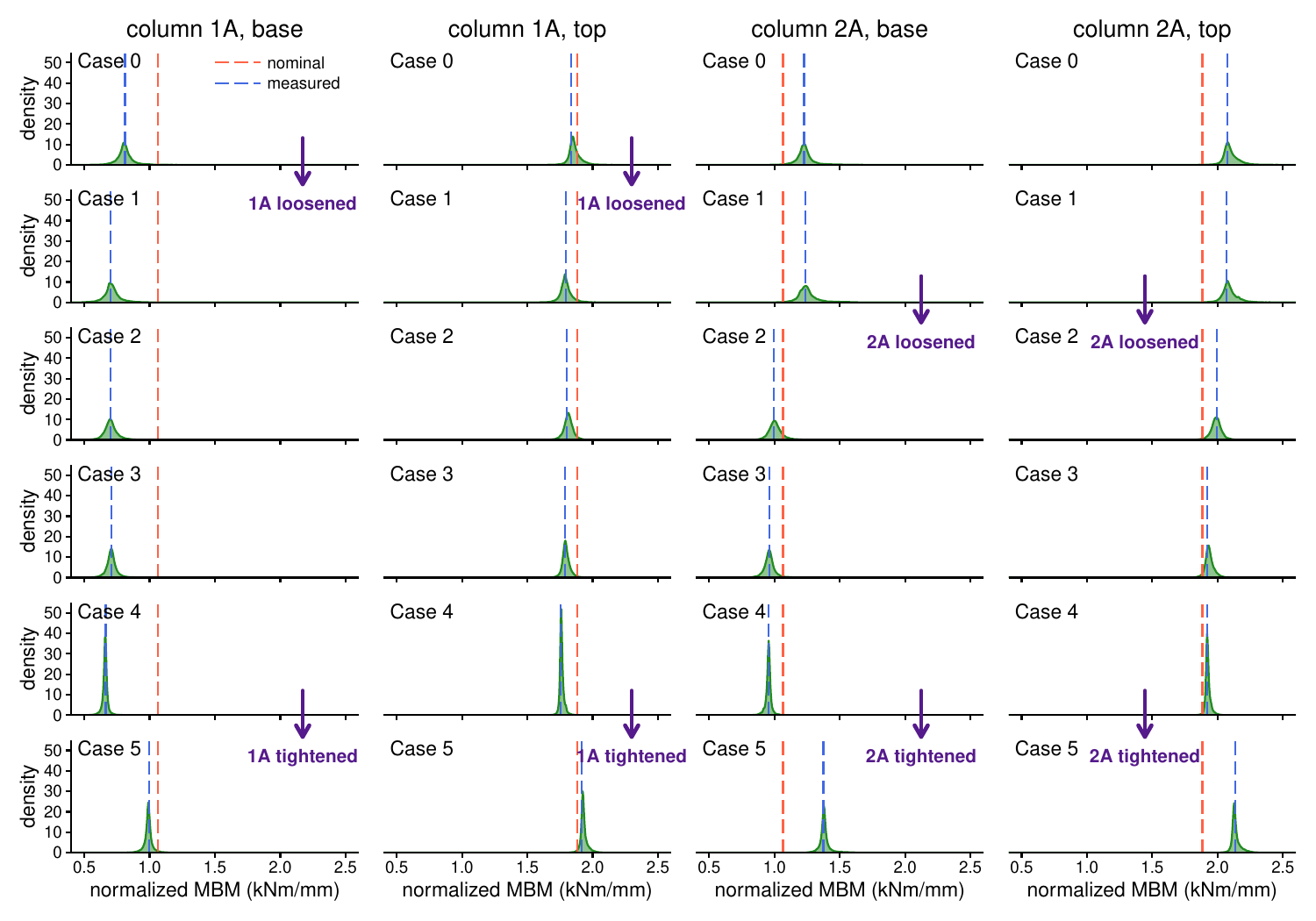}
    \caption{Predictive distributions of normalized modal bending moments at different sections in Frame A with the nominal and measured values.}\label{fig:density}
\end{figure}

Figure~\ref{fig:kbs} shows the posterior distributions of the rotational stiffness of all column bases for different cases and directions for both SMT and IST scenarios.
For example, the values for column 1A correspond to ${}_\mathrm{A}k_1$ (in the EW direction) and ${}_1k_1$ (in the NS direction). The gray area represents the cases in which the anchor bolts of the column base are loosened. For example, the anchor bolts of column 1A are loosened in Cases 1--4.
For both scenarios, loosening the anchor bolts significantly decreased the rotational stiffness for all column bases.
For example, the 0.05-quantile values in Cases 3 and 4 (with loosened anchor bolts) were below the 0.95-quantile values in Cases 0--2 (with tightened anchor bolts for column 1B in both directions).
Similarly, re-tightening anchor bolts significantly increased the rotational stiffness of all column bases from Cases 4 to 5.
These results suggest that the proposed method successfully localized the stiffness changes in column bases with a limited degree of uncertainty.
The green horizontal line represents nominal values given by Eq.~(\ref{eq:nominal}), which are of the same order of magnitude as the identified values, thereby validating the results.
However, a significant difference exists between the nominal value and the identified stiffness, even in the initial state. For example, the identified values for column 1A are almost half of the nominal value.
This implies the necessity of updating the model to reflect the characteristics of different column bases.

Figure~\ref{fig:bmf} shows the posterior distributions of the beam rigidification coefficients for different frames transformed into $10^{{}_f\gamma} ~ (= {}_f {EI}_\mathrm{cs} / {}_f EI_\mathrm{s})$, comparing the two inference scenarios.
For all frames, the (a) SMT scenario shows that the width of the distribution gradually decreases from Cases 0 to 5, whereas the (b) IST scenario shows a larger width in all cases. This suggests that using sequentially obtained measurements can significantly reduce uncertainties.
Even in Case 0 with no prior measurement, the (a) SMT scenario exhibits a narrower width between 0.25 and 0.75 quantiles and closer medians across different frames, compared to the (b) IST scenario.
This suggests that hierarchical multitask modeling allows for more stable and reasonable inferences.

Figure~\ref{fig:density} shows the predictive distributions of the normalized MBMs at four sections in Frame A, i.e., the base and top of columns 1A and 2A, computed using the posterior samples of the parameters ${}_fk_1, {}_fk_2, {}_f\gamma$ for the SMT scenario with the measured values indicated by the blue dashed lines. The red dashed lines represent the nominal normalized MBM values computed using the nominal values of the rotational stiffness and beam rigidification coefficients. 
In Case 0, the nominal values revealed large discrepancies from the measurements in the intact state, especially at the base of column 1A.
This suggests the importance of updating the model for performance evaluation of existing structures.
In contrast, peaks of the predictive distributions were considerably closer to the measurements for all sections in all cases. The distributions successfully captured the effect of loosening column base 1A on the base and the top of the column from Cases 0 to 1, that of loosening column base 2A on the base and top of the column from Cases 1 to 2, and that of re-tightening all column bases on all sections from Cases 4 to 5.
Furthermore, the narrower and sharper distributions in Cases 0 to 5 reflect the reduced uncertainty of the beam rigidification observed in Figure~\ref{fig:bmf}. This implies the effectiveness of multitask modeling that transfers knowledge across different frames and cases in the SMT scenario.

\section{Conclusions}

A multitask learning framework for probabilistic model updating jointly using strain and acceleration measurements was devised in this study. The framework can yield enhanced structural damage assessment and response predictions of existing steel structures with quantified uncertainty. This method includes three components: (1) system identification for modal parameter estimation; (2) evaluation of the displacement--stress relationship in the modal space; and (3) updating of the model by using this relationship based on Bayesian multitask learning to help different tasks (multiple planar frames constituting the spatial structure of interest) transfer knowledge of the model parameters to each other.
The proposed method was validated using a full-scale vibration test on a moment-resisting steel frame structure wherein the structural damage to the column bases was simulated by loosening the anchor bolts.
This experimental results indicate that the proposed metric of normalized MBMs has sufficient sensitivity toward localized damage, for example, a decrease in the rotational stiffness of a column base. Therefore, the normalized-MBM-based model updating, which uses both strain and acceleration measurements, facilitates sensitive damage localization and quantification in steel frame structures. Moreover, the Bayesian multitask learning approach that was utilized to achieve more efficient use of the measurement data can reduce the uncertainty involved in model parameter estimation and allow for more robust and informative model updating.

As regards future research directions, first, application of the proposed method to seismic responses, as discussed in 3.1, should be addressed using modal parameters estimated via MIMO system identification.
Second, an extended formulation should be investigated to ensure the applicability of the proposed framework to multistory multibay frame structures. This formulation should consider multiple damage scenarios including damage to beam-column connections.

\section*{Acknowledgements}

The dynamic loading test in this study was a part of the research conducted in collaboration with Prof. Satoshi Yamada and Prof. Tsuyoshi Seike at the University of Tokyo and supported by JSPS KAKENHI
(Grant-in-Aid for Scientific Research) (B) Grant Number JP20H02293.

%% \appendix
%% \section{}
%% \label{}

\bibliographystyle{elsarticle-num} 
\bibliography{bib_revised}

\end{document}